\documentstyle[prl,aps,epsfig,amssymb,multicol]{revtex}
\def\hw{$\hbar\omega\,$}
\def\H{${\cal H}$}
\def\Hm{${\cal H}_m$}
\def\Hmd{${\cal H}_m^d$}
\def\Hmtd{${\cal H}_{mT}^d$}
\def\HM{${\cal H}_M$}
\begin{document}
 \draft
\title{Monopole, quadrupole and pairing: a shell model view} 
\author{ A. P. Zuker}
\address{ IRES, B\^at27, IN2P3-CNRS/Universit\'e Louis
Pasteur BP 28, F-67037 Strasbourg Cedex 2, France}
\date{\today}
\maketitle
\begin{abstract}
The three main contributions to the nuclear Hamiltonian---monopole,
quadrupole and pairing---are analyzed in a shell model context. The
first has to be treated phenomenologically, while the other two can be
reliably extracted from the realistic interactions. Due to simple
scaling properties, the realistic quadrupole and pairing interactions
eliminate the tendency to collapse of their conventional counterparts,
while retaining their basic simplicity.
\end{abstract}
\begin{multicols}{2}
\narrowtext
Nuclei are systems of interacting particles. Everybody agrees with
this statement. But then: which interaction, which particles? These
questions raise some difficulties, so the first statement is replaced
by : Nuclei are systems of quasi-particles, interacting via effective
Hamiltonians (or Lagrangians). The questions recur, and now the
answers vary widely. Restricting attention to theorists who are content
with the idea that the particles are basically neutrons and protons,
there main possibilities concerning the interaction:

{\bf R}. The realistic approach consists in extracting it from $NN$ data,
and then take seriously the idea that the many-body Sch\"{o}dinger has
to be solved exactly.

{\bf P}. The phenomenological approach(es) derive the interaction from the
nuclear data they are supposed to explain. There are two variants:
\begin{description}
\item [MFP] Mean field phenomenology restricts to a maximumn the
  number of parameters, and directs attention to global properties and
  general spectroscopic trends.
\item [SMP] Shell model phenomenology is prepared to introduce as many
  parameters as necessary to explain spectroscopic detail.
\end{description}

Conceptually, {\bf R} is the most satisfactory but, in addition to its
difficulty, it has suffered so far from a major drawback: The
realistic interactions do not saturate well, i. e., they do not give
the right binding at the right radius. Furthermore, they fail to
produce the correct shell structure: doubly magic closures are missed.
Much progress has been made recently in fitting the $NN$ data
perfectly, and in calculating exactly for very light nuclei. However,
the problems remain, and phenomenological three-body forces have to be
called in to solve them.

Historically, {\bf P}, has enormous importance, and both its variants
rest on pioneering work that remains of permanent value. It can be
summarized by saying that, whatever their guise, the effective
Hamiltonians must contain a monopole term that produces the spherical
closures, a quadrupole one that induces deformation, and a pairing
force that favours condensation at the vicinity of the Fermi surface.

My purpose is to show how these things are possible, {\em within a
  framework that starts from} {\bf R}. The idea is that the trouble in
realistic interactions is concentrated in the monopole part, which can
be separated rigorously from the rest. Moreover, it is formally very
simple and can be extracted from the data with a mimimum of
parameters. Once this is done, the rest of the Hamiltonian (which we
call multipole), turns out to be independent of the type of realistic
interaction used. What's more, it does a magnificent job in large
scale shell model calculations.

The aim of these notes~\cite{foot} is to provide a unified introduction to
references~\cite{duf99,duf96}, dealing with the monopole and multipole
Hamiltonians respectively.
 
\smallskip

{\em NOTATIONS} A few equations have to exhibit explicitly angular
momentum ($J$), and isospin ($T$) conservation. I will use Bruce
French's product notation~\cite{fre66})

$\Gamma$ stands for $JT$. Then $(-)^\Gamma=(-)^{J+T}$,
$[\Gamma]=(2J+1)(2T+1)$, and in general $F(\Gamma)=F(J)F(T)$.
Orbits are called $r$, $s$, etc., and I use $(-)^r=(-)^{j_r+1/2}$,
$[r]=2(2j_r+1)$. Expressions carry to neutron-proton formalism simply
by dropping the issospin factor

$m_r$ is the number of particles in orbit $r$, $T_r$ is used for both
the isospin and the isospin operator. In neutron-proton ($np$) scheme,
$m_{rx}$ specifies the fluid $x$.

$Z^{\dagger}_{rs\Gamma}$ is an operator of type $a^{\dagger}_r
a^{\dagger}_s$ coupled to good spin and isospin $JT$. $S^\gamma _{rs}$
is an operator of type $a^{\dagger}_r a_s$ coupled to good spin and
isospin $\lambda \tau$.

$V^{\Gamma}_{rstu}$ is a two body matrix element. $W^{\Gamma}_{rstu}$
is used after the monopole part has been subtracted.

$p$ is the principal oscillator quantum number.

\section{The monopole Hamiltonian}
\label{sec:m}
The only assumption that will be made is the existence of an effective
potential smooth enough to do Hartee Fock (HF) variation, and
 capable of yielding good results. 

Now: given a
Hamiltonian \H, (${\cal K}$ is the kinetic energy)

  \begin{equation}
{\cal H}={\cal K}+ \sum\limits_{r\leq s,t\leq u,\Gamma}
V_{rstu}^\Gamma Z_{rs\Gamma}^+
\cdot Z_{tu\Gamma},\label{H}
\end{equation}

 it is always possible to extract from it
a monopole part \Hmd, whose expectation value for any state is the
average energy of the configuration to which it belongs (a
configuration is a set of states with fixed $m_{rx}$ for each orbit).
In particular, \Hmd reproduces the exact energy of closed shells ($cs$)
and single particle (or hole) states built on them ($(cs)\pm 1$),
since for this set ($cs\pm 1$) each configuration contains a single
member. The result is standard~\cite{fre69}, and we simply write it,
($d$ stands for diagonal)

\begin{equation}
\label{hmnp}
{\cal H}_m^d={\cal K}^d+ \sum_{rx,sx'}
V^{xx'}_{rs}m_{rx}(m_{sx'}-{\delta}_{rs}{\delta}_{xx'}),
\end{equation}
which reproduces the average energies of configurations at fixed
$m_{rx}m_{rx'}$.  and isospin ($mT$)
\begin{eqnarray}
&&{\cal H}_{mT}^d ={\cal K}^d+\sum_{r\leq s}{1 \over (1+\delta_{rs})}
\bigl[a_{rs}\,m_r(m_s-\delta_{rs})+\nonumber\\
&&             +b_{rs}(T_r\cdot  T_s-{3\over 4}m\delta_{rs})\bigr],
\label{hmt}
\end{eqnarray}
which reproduces the average energies of configurations at fixed $m_rT_r$.

Using $D_r=2j_r+1$, we rewrite the
relevant centroids incorporating explicitly the Pauli restrictions
\begin{eqnarray}
V_{rx,sx'}&=&{\sum_J V_{rsrs}^{Jxx'}(2J+1)
  ( 1-(-)^J\delta_{rs}\delta_{xx'})\over
  D_r(D_s-\delta_{rs}\delta_{xx'})}\label{vnp}\\
V_{rs}^T&=&{\sum_J V_{rsrs}^{JT}(2J+1)
( 1-(-)^{J+T}\delta_{rs})\over D_r(D_s+\delta_{rs}(-)^T)}
\label{vjt}\\
a_{rs}&=&{1\over 4}(3V^1_{rs}+V^0_{rs}), \quad
b_{rs}=V^1_{rs}-V^0_{rs}.
\label{ab}
\end{eqnarray}

In the $np$ scheme each orbit $r$ goes into two $rx$ and $rx'$
and the centroids can be obtained through
 ($x\ne x'$)
\begin{eqnarray}
V_{rx,sx'}&=&{1\over 2}\left\lbrack V_{rs}^1\left(1-{\delta_{rs}\over D_r}
\right)+ V_{rs}^0\left(1+{\delta_{rs}\over D_r}
\right)\right\rbrack\nonumber\\
V_{rx,sx}&=&V_{rs}^1.
\label{npmt}
\end{eqnarray}

 Under Hartree Fock variation, the $m_r$ and $T_r$
 operators will go into non diagonal ones of the type $S^{0\tau}_{st}$.
 Therefore \Hmd and \Hmtd should be generalized to the full monopole
 Hamiltonian, ${\cal H}_m$, containing  all
 two body quadratic forms in the $S^{0\tau}_{st}$ operators,
 see~\cite{zuk94,zuk95}). The task is not trivial, and we can avoid
 doing the variation explicitly, as we see next.

\subsection{Scaling}
\label{sec:s}
HF variation is necessary to ensure that the system assume its correct
radius, at an energy close to the correct one. In all representations
\Hmd has the same form, that must change smoothly as $N$ and $Z$
change. To discover how the evolution takes place, we first note that
matrix elements for a potential of short---but non zero---range scale
as the oscillator frequency \hw (a consequence of the Talmi Moshinsky
transformation). Then we can write

\begin{equation}
V(\omega)_{klmn}\cong {\omega\over{\omega}_0}
  V({\omega}_0)_{klmn}, \quad \hbar \omega=34.6A^{1/3}/\langle r^{2},
\label{eq:scal}   
\end{equation}
where the second equality is adapted from~\cite{BM64}. A very accurate
fit to the spherical radii (a variant of
those in~\cite{duf94}) yields $\langle r^2\rangle =0.89\rho A^{1/3}$ with 
\begin{equation} 
\label{rho}
\rho=A^{1/3}(1-(2T/A)^2)e^{(3.5/A)}, \; {\rm so}\quad \hbar
\omega\approx 39/\rho.  
\end{equation}

This is very simple, but it cannot be the full story, because the
competition between potential and kinetic energy (which has the same
scaling, \mbox{${\cal K}^d\equiv K=\hbar\omega/2\sum_p {(p+3/2)m_p}$},
$m_p$ is the number of particles in harmonic oscillator (HO) shell
$p$) would lead to a trivial equilibrium at $\hbar \omega=\infty$ or
0. The terms that cannot scale strictly with $\omega$ must be those
that go as the total number of particles $A$, since they are alone
responsible for saturation. We shall not try to discover the
saturation mechanism, but simply note that \Hmd should produce (the
right) smooth contributions in $A$ and $A^{2/3}$, and a symmetry
energy in $T(T+1)/A$ and $T(T+1)/A^{4/3}$. Next we separate these
``liquid drop'' terms, from those that produce shell effects
\Hmd=${\cal H}_m^{LD}+{\cal H}_m^s$, and set out to parametrize
efficiently ${\cal H}_m^s$.

\subsubsection{${\cal H}_m^s$ and the collective monoplole Hamiltonian $W$}
\label{sec:W}

By {\em definition}, shell effects go as $A^{1/3}$, and we have
assumed they can be separated those in $A$ and $A^{2/3}$. Strutinsky's
famous theorem tells us this is possible. Here I will show an explicit
example, which is the starting point in the construction of ${\cal
  H}_m^s$. Independently of the detailed saturation mechanism, there
must be ``something'' in ${\cal H}_m^d$ capable of cancelling $K$. As
it is always possible to separate from ${\cal H}_m^d$ a leading term
in $m(m-1)/2\equiv A(A-1)/2$, the ``something'' must be related to it.
As the end result must go like $A$, this term must be scaled by
$A^{-1}$. However, there is no such thing in ${\cal H}_m^d$ if we take
the scaling to be \hw, i. e., $A^{-1/3}$. By using the techniques of
section~\ref{sec:M}, we discover that the term we are after is

\begin{equation}
  \label{W}
 W=\sum_p{(m_p/\sqrt{D_p})^2}, \quad  D_p=(p+1)(p+2),
\end{equation}
which has the remarkable property that $\hbar \omega (W-4K)/4$ cancels
exactly to order $A^{1/3}$ and (as we shall see) produces strong shell
effects. $W$ is called the collective monopole Hamiltonian for reasons
that will become fully apparent in section~\ref{sec:c}.

\begin{figure}[htb]
      \begin{center}
      \leavevmode
      \epsfig{file=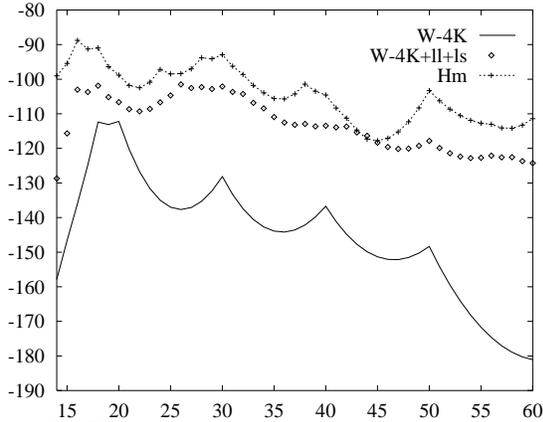,width=8cm}
      \caption{Shell formation mechanism in $T=5$ nuclei. Negative
        binding energies in $y$-axis are {\em repulsive}. $N$ in
        $x$-axis}.
      \label{fig:hmt10}
    \end{center}
 \end{figure}

$W-4K$ is taken to be the first contibution to ${\cal H}_m^d$, which
has been fitted on the set  of known $(cs)\pm 1$ spectra.  For the
explicit construction of the rest I refer to~\cite{duf99}. For the
present purpose it will be sufficient to mention that it includes
an  $l\cdot s+l\cdot l$ piece that behaves as one-body and has the
same coefficient as $W-4K$. Then, there are two-body terms that ensure
the correct evolution of $(cs)\pm 1$ spacings from HO to EI closures
(extruder-intruder, such as $N\, Z=14$, 28, 50,
etc). Fig~\ref{fig:hmt10} illustrates the shell formation mechanism
for $T=5$ nuclei. Note the beautiful HO closures generated by $W-4K$,
erased by $l\cdot s+l\cdot l$ (as in $^{30}$Ne at $N=20$), and
replaced by EI closures through the two body terms.

\section{The multipole Hamiltonian}
\label{sec:M}
The multipole Hamiltonian is defined as \HM=\H-\Hm. As we are no
longer interested in the full \H, but its restriction to a finite
space, \HM will be more modestly called $H_M$, with monopole-free
matrix elements given by
\begin{equation}
  \label{VW}
  W_{rstu}^{JT}=V_{rstu}^{JT}-\delta_{rs}\delta_{tu}V_{rs}^T.
\end{equation}
To concentrate on the physics I shall only write the formulae that are
strictly necessary to show how the pairing plus quadrupole
interactions appear in their {\em normalized} form of. Abundant detail
is given in~\cite{duf96}.

There are two standard ways of writing $H_M$:

 \begin{equation}
 H_M=\sum\limits_{r\leq s,t\leq u,\Gamma}
W_{rstu}^\Gamma Z_{rs\Gamma}^+
\cdot Z_{tu\Gamma},\quad {\rm or}
\label{1}
\end{equation}
\begin{equation}
 H_M=\sum_{rstu \Gamma}[\gamma]^{1/2}\omega_{rtsu}^\gamma (S_{rt}^\gamma
S_{su}^\gamma)^0,
\label{2}
\end{equation}
where $\zeta_{rs}=(1+\delta_{rs})^{1/2}/2$ , and the matrix elements are
related through
\begin{equation}
\omega_{rtsu}^\gamma =\sum_\Gamma(-)^{s+t-\gamma-\Gamma}
\left\{\begin{array}{ccc}
   r&s&\Gamma\\  u&t&\gamma
   \end{array}\right\}
W_{rstu}^\Gamma[\Gamma],\label{3a}
\end{equation}
\begin{equation}
W_{rstu}^\Gamma =\sum_\gamma(-)^{s+t-\gamma-\Gamma}
\left\{\begin{array}{ccc}
   r&s&\Gamma\\  u&t&\gamma
   \end{array}\right\}
\omega_{rtsu}^\gamma[\gamma].\label{3b}
\end{equation}

Replacing pairs by single indeces $rs\equiv x$, $tu=y$ in
eq.~(\ref{1}) and $rt\equiv a$, $su=b$ in eq.~(\ref{2}), we bring the
matrices $W_{xy}^\Gamma$ and $f_{ab}^\gamma\equiv
f_{rtsu}=\zeta_{rs}\zeta_{tu}$, to diagonal form through unitary
transformations $U_{xk}^\Gamma,u_{ak}^\gamma$:

\begin{equation}
U^{-1}WU=E  \Longrightarrow W_{xy}^\Gamma =\sum_k
U_{xk}^\Gamma U_{yk}^\Gamma E_k^\Gamma \label{III.3}
\end{equation}
\begin{equation}
u^{-1}fu=e\Longrightarrow f_{ab}^\gamma =\sum_k
u_{ak}^\gamma u_{bk}^\gamma e_k^\gamma, \label{III.4}
\end{equation}
and then,

\begin{equation}
H_M=\sum_{k,\Gamma}E_k^\Gamma\sum_xU_{xk}^\Gamma
Z_{x\Gamma}^+\cdot\sum\limits_yU_{yk}^\Gamma Z_{y\Gamma},\label{4a}
\end{equation}
\begin{equation}
H_M=\sum_{k,\gamma}e_k^\gamma\left(\sum_au_{ak}^\gamma
S_a^\gamma\sum_bu_{bk}^\gamma S_b^\gamma\right)^0
[\gamma]^{1/2},\label{4b}
\end{equation}
which we call the $E$ and $e$ representations.  Since ${\cal H}_m$
contains all the $\gamma=00$ and 01 terms, for ${\cal H}_M$,
$\omega_{rstu}^{00}=\omega_{rstu}^{01}=0$.  There no one body
contractions in the $e$ representation because they are all
proportional to $\omega_{rstu}^{0\tau}$.

\begin{figure}
  \begin{center}
    \leavevmode
    \psfig{file=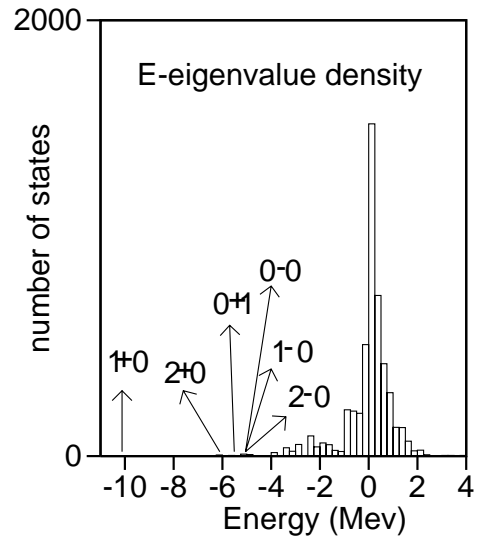,angle=270,width=8cm}
\vspace {1.cm}
\caption{$E$-eigenvalue density for the KLS interaction in the pf+sdg
major shells $\hbar\omega=9$. Each eigenvalue has multiplicity $[\Gamma]$.
The largest ones are shown by arrows.}\label{Edef}
  \end{center}
\end{figure}

The eigensolutions in eqs.~(\ref{4a}) and (\ref{4b}) using the KLS
interaction~\cite{KLS69,LEE69}---in ref.~\cite{duf96} it is explained
in detail why this venerable choice is a good one---for spaces of one
and two major oscillator shells. The density of
eigenvalues (their number in a given interval) in the $E$
representation is shown in fig.~\ref{Edef} for a typical two-shell case.

It is skewed, with a tail at negative energies which is what we expect
from an attractive interaction.
\begin{figure}
  \begin{center}
    \leavevmode
    \psfig{file=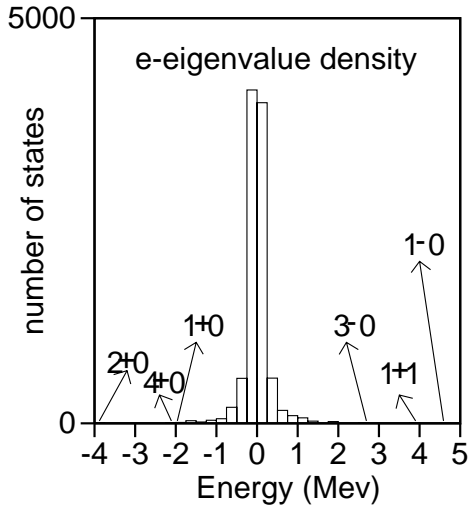,angle=270,width=8cm}
\vspace {1.cm}
\caption{$e$-eigenvalue density for the KLS interaction in the pf+sdg
major shells. Each eigenvalue has multiplicity $[\gamma]$.
The largest ones are shown by arrows.}\label{edef}
  \end{center}
\end{figure}

\begin{figure}
  \begin{center}
    \leavevmode
    \psfig{file=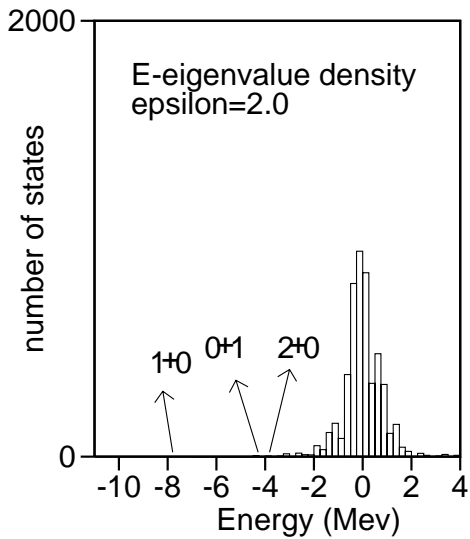,angle=270,width=8cm}
\vspace {1.cm}
\caption{$E$-eigenvalue density for the KLS interaction in the pf+sdg
major shells $\hbar\omega=9$, after removal of the five largest multipole
contributions. Each eigenvalue has multiplicity $[\Gamma]$.
The largest ones are shown by arrows.}\label{Edef20}
  \end{center}
\end{figure}
The $e$ eigenvalues have a number of simple properties demonstrated in
\cite [Appendix B]{duf96} : their mean value always vanishes, their
width is $\sqrt{1/8}$ of that of the $E$ distribution, and they are
twice as numerous.  In fig.~\ref{edef} we find that they are very
symmetrically distributed around a narrow central group, but few of
them are neatly detached. The strongest have
$\gamma^\pi=1^-0,\;1^+1,~2^+0,~3^-0, ~4^+0$. {\bf If the corresponding
  eigenvectors are eliminated from H in} eq.~(\ref{4b}) {\it and the
  associated H in} eq.~(\ref{4a}) {\it is recalculated, the E
  distribution becomes quite symmetric}. (The boldface is meant to call
attention to the one very bad erratum in~\cite{duf96}). The result is
shown in Fig.~\ref{Edef20}. The residual
skewness is entirely accounted for by the $\Gamma=1^+0,\; 0^+1
\;$and$\; 2^+0$ peaks, whose strength is somewhat eroded but remains
substantial.

This result is most telling because from the work of Mon and French
\cite{mon75} (see also the references quoted in~\cite{foot}) we know
that a symmetric $E$ distribution will lead to spectra in the
$m$-particle systems that are identical to those of a random matrix.
Therefore, we have found that - with the exception of three $\Gamma$
peaks - the very few dominant terms in the $e$-distribution are
responsible for deviations from random behaviour in $H_M$.  Positively
stated, these terms are at the origin of collective properties.

If the diagonalizations are restricted to one major shell, negative
parity peaks are absent, but for the positive parity ones the results
are practically identical to those of Figs.~\ref{Edef} and \ref{edef},
except that the energies are halved. This point is crucial: 

{\em If $u_{p_1}$ and $u_{p_2}$ are the eigenvectors obtained in shells
  $p_1$ and $p_2$, their eigenvalues are approximately equal
  $e_{p_1}\approx e_{p_2}=e$. 

When diagonalizing in $p_1+p_2$, the {\em unnormalized}
  eigenvector turns out to be $u_{p_1}+u_{p_2}$ with eigenvalue $e$}. 

In the figures the eigenvalues for the two shell case are doubled,
because they are associated with normalized eigenvectors.

To make a long story short: The contribution to $H_M$ associated to the
$\Gamma=01$, and $\gamma=20$,

\begin{eqnarray}
H_{\bar P}&=&-{\hbar\omega\over\hbar\omega_0}{|E^{01}|}
(\overline P\,^+_p+\overline P\,^+_{p+1})
\cdot(\overline P_p+\overline P_{p+1})\label{Hp}\\
H_{\bar q}&=&-{\hbar\omega\over\hbar\omega_0}{|e^{20}|}
(\bar q_p+\bar q_{p+1})
\cdot(\bar q_p+\bar q_{p+1}),\label{Hq}
\end{eqnarray}
turn out to be (naturally)
the usual pairing plus quadrupole Hamiltonians, {\em except that
the operators for each major shell of principal quantum number $p$ are
affected by a normalization}. $E^{01}$ and $e^{20}$ are the {\em one}
shell values called generically $e$ in the discussion above. To be precise

\begin{eqnarray}
\bar P^+_p&=&
\sum\limits_{r\in p}Z_{rr01}^+\Omega_r^{1/2}/\Omega_p^{1/2},\label{P}\\
\bar q_p&=&
\sum\limits_{rs\in p}S_{rs}^{20}q_{rs}/{\cal N}_p,\label{q}
\end{eqnarray}
where

$\Omega_r=j_r+1/2$, $q_{rs}=\sqrt{1\over5}\langle\| r^2Y^2\|
s\rangle$, and

\begin{equation} \label{norms}
\Omega_p=\sum_{r}\Omega_r={1\over 2}D_p \quad
{\cal N}_p^2=\Sigma q_{rs}^2\cong {5\over 32\pi}(p+3/2)^4,
\end{equation}

More on the other collective terms of $H_M$ in section~\ref{sec:mi}

\subsection{Collapse avoided}\label{sec:c}

The pairing plus quadrupole model has a long and glorious
history~\cite{bar68,bes69}, and one big problem: as more shells are
added to a space, the energy grows, eventually leading to
collapse. The only solution is to stay within limited spaces, but then
the coupling constants have to be readjusted on a case by case
basis. The normalized versions of the operators presented above are
affected by universal coupling constants that dot change with the
number of shells. Knowing that \hw$_0=9$ MeV, they are $|E^{01}|/{\hbar
  \omega_0}=g'=0.32$ and $|e^{20}|/{\hbar \omega_0}=\kappa '=0.216$ in
Eqs.~\ref{Hp} and \ref{Hq}. (One should not forget that for practical
use, these numbers must be renormalized. See~\cite{duf96} for details.)  
 
Introducing $A_{mf}\approx{2\over 3}(p_f+3/2)^3$, the total number of
particles at the middle of the Fermi shell $p_f$, the relationship
between $g'$, $\kappa '$, and their conventional
counterparts~\cite{bar68} is, for one shell 

\begin{eqnarray}
{0.32\hbar\omega\over \Omega_p}&\cong&
{19.51\over A^{1/3}A_{mf}^{2/3}}= G \equiv {G_0}A^{-1},
\nonumber\\
{0.216\hbar\omega\over {{\cal N}_p^2}}&\cong& {1\over 2}\,
{216\over A^{1/3}A_{mf}^{4/3}}={\chi'\over 2}\equiv 
{\chi_0'\over 2}A^{-5/3}.
\label{1s}
\end{eqnarray}

What I propose next, is to show that both variants do indeed produce
shell effects in $A^{1/3}$, a necessity emphasized at the beginning of
section~\ref{sec:W}. Then, by allowing the particles to be promoted to
shells well above the Fermi one, it will become clear why the
conventional forms produce collapse, and the new ones do not.

Assume $m=O(D_f)$ particles in shell $p_f$, for which, $D_f=O(A^{2/3})$.

Consider first the pairing force. On a space of degeneracy $D$, it
produces an energy
\[E_P=-{|G|\over 4}m(D-m+2)=-|G|O(mD).\]
The first equality is a standard result. For $D=D_f$, the
conventional choice $G=O(A^{-1})$ leads to $E_P=O(A^{1/3})$, i. e., it
guarantees the correct scaling for shell effects. That the result is
not as trivial as it looks can be gathered from this quotation
concerning the $G$ coupling: ``We know of no reliable way of predicting
this $A^{-1}$ dependence...''\cite{bar68}.

For a quadrupole force, an estimate for the energy can be obtained by
constructing a determinantal state that maximizes the quadrupole
moment \mbox{$Q_0=\sum_{i=1}^{n}(2n_{zi}-n_{xi}-n_{yi})$}, where
$n_{xi},n_{yi},n_{zi}$ are the number of quanta. The largest term in
the sum is then $2p$, the next $2p-3$, then $2p-6$, etc. Therefore
$Q_0=O(mp)$, and
\[E_q\approx -|\chi'|Q_0^2=-|\chi'|O(m^2D),\]
which in turn explains (for $D=D_f$) the origin of the usual choice
$\chi'=O(A^{-5/3})$ for the quadrupole strength, that leads to
$E_q=O(A^{1/3})$.

It is clear from equations (\ref{1s}) that the
operators are affected by coefficients that go as $A^{-1/3}D^{-1}$
(instead of $A^{-1}$) for pairing, and as $A^{-1/3}D^{-2}$ (instead of
$A^{-5/3}$) for quadrupole. For $D=D_f$, the energies are again
$O(A^{1/3})$, but now this important empirical fact is a direct
consequence of the interaction. For arbitrary $D$, the energies of the
traditional (old) versions transform into the normalized (new) one as

\[E_P({\rm old})=O({mD\over A})\Longrightarrow E_P({\rm
  new})=O({m\over A^{1/3}}) 
,\]

\[ E_q({\rm old})=O({m^2D\over A^{5/3}})\Longrightarrow  E_q({\rm
  new})=O({m^2\over A^{1/3}D}).\]

If the m particles are promoted to some higher shell with $p=p_f+M$,
$D\approx (p_f+M)^2$, both energies grow in the old version.
For sufficiently large $M$ - because of the term in $M^2$ - the gain
will become larger than the monopole loss
$O(mM\hbar\omega)=O(MA^{1/3})$ that is only linear in $M$.  Therefore
the traditional forces lead the system to collapse. In the new form
there is no collapse: $E_P$ stays constant, $E_q$ decreases and the
monopole term provides the restoring force that guarantees that
particles will remain predominantly in the Fermi shell.

Let us go back now to the monopole term $W$ that plays such a crucial
role in \Hmd. The collective operators are obviously $m_p$, and the
``conventional'' monopole force $V_0mm=V_0(\sum m_p)^2$. Granted that
$V_0$ would be scaled by $A^{-1}$, as a centroid, it would have to be
calculated for the occupied orbits only, which would have to be varied
to detect a minumum. In other words: there is no collapse, but the
operator is useless: All the information is contained in $V_0$, which
is not a constant, but the result of HF variation from nucleus to
nucleus.

However, it is clear that $W$ in Eq.~(\ref{W}) is the normalized
version of the $mm$ operator, and that it is quite a useful.

\section{Pairing miscellanea}\label{sec:mi}
The treatment of the monopole term is new, and I can only hope that
the techniques and results in \cite{duf99} will stimulate further
interst in \Hm. 

The quadrupole force is very old, and nobody seems to have much
problems concerning it. Here I only note that the quadrupole terms
involving $2\hbar \omega$ jumps are quite suppressed in the realistic
force: They happen to be exactly what is needed to produce the correct
effective charges in perturbation theory.  Therefore we are left
basically with a $q\cdot q$ force of Elliott's type, associated with
the SU(3) symmetry~\cite{ell56}.

The other big multipole terms in the collective Hamiltonian are very
much what they should be: octupole, hexadecapole and $\sigma \tau$
forces. There is also a huge $\gamma=10^-$ center of mass
contribution, that is nicely decoupled, as it should, to ensure
momentum conservation.

We are left with the pairing terms. Contrary to quadrupole---which
seems to be naturally and satisfactorily included in mean field
formulations---pairing has to be put by hand. Since the old version is
not sufficient for state of the art calculations, there are many
proposals to replace it. Furthermore, there is also much interest in
the other pairing terms. Let us start with a look at these.

The very large $\Gamma=10$ one is the $ST=10$ part of a
pairing force in $LS$ scheme ($\Gamma=01$ is indeed the $ST=01$
part). In $jj$ coupling, the term is massively dominated by the matrix
elements between the largest $l\cdot s$ partners (e. g.,
$f_{7/2}f_{5/2}$ in $p=3$). In spite of its strength it is much
suppressed by the splitting between the orbits. Furthermore, its $T=0$
nature makes it rapidly inefficient when moving away of $T=0$ nuclei.
It is also at the origin of a common misconception regarding the
Wigner term, that deserves a digression.

{\bf The Wigner term} is the piece of the force that goes like $T$. It
simply comes from the $T(T+1)$ symmetry energy which is very strong,
and produces a cusp in the binding energies at $T=0$. Therefore,
before deciding that $\Gamma=10$ (or some $J0$) pairing is responsible
for the Wigner term, people are asked to check that the coefficient of
the symmetry energy (a centroid of type $b$ in Eq.~(\ref{ab})) has
been kept constant.

The $\Gamma=20$ term is a puzzle. I do not know what to make of it.

There are no other pairing terms that may claim a collective status,
in particular, the $\Gamma=21$ term does not seem to amount to much. I
would strongly urge people to stop using it.

Finally, let me return to $\Gamma=01$ pairing. As mentioned, there are
various new candidates. Once the ones that do not make sense are
discarded (I have in mind the $\delta$ force), my feeling is that
something like the version that has been proposed here will be either
accepted or independently discovered.  Modifications may be necessary,
because there are problems (discussed in~\cite{duf96}), but the
proposal seems quite sound.

Refs.~\cite{CEM*95,zuk97,pov98} contain discussions about the
influence of $\Gamma=01,\, 10$ pairing on backbending rotors, that can
be summed up as follows:

{\em Although the energetics of the yrast band are strongly affected
by the pairing modifications, the other properties are not, since the
wavefunctions change little}.

Nuclear physics is not confined to backbending rotors. In other
regions, pairing will dominate the coupling schemes, with quadrupole
acting as a pertubation. In general, it seems quite likely that
nuclear structure will remain a field in which pairing and quadrupole
are the main players, with monopole acting as referee.

\end{multicols}
\end{document}